\documentclass[a4paper,english,aps,prb,showpacs,floatfix,nofootinbib,twocolumn]{revtex4}
\usepackage{mathptmx}
\usepackage[T1]{fontenc}
\usepackage[latin9]{inputenc}
\usepackage{color}
\usepackage{babel}
\usepackage{amsmath}
\usepackage{amssymb}
\usepackage{graphicx}
\usepackage{esint}
\usepackage[unicode=true,
 bookmarks=true,bookmarksnumbered=true,bookmarksopen=false,
 breaklinks=true,pdfborder={0 0 0},backref=false,colorlinks=true]
 {hyperref}
\hypersetup{pdftitle={Graphene Paper},
 pdfauthor={Vasilios D. Karanikolas}}

\makeatletter

\pdfpageheight\paperheight
\pdfpagewidth\paperwidth

\providecommand{\tabularnewline}{\\}

\@ifundefined{textcolor}{}
{%
 \definecolor{BLACK}{gray}{0}
 \definecolor{WHITE}{gray}{1}
 \definecolor{RED}{rgb}{1,0,0}
 \definecolor{GREEN}{rgb}{0,1,0}
 \definecolor{BLUE}{rgb}{0,0,1}
 \definecolor{CYAN}{cmyk}{1,0,0,0}
 \definecolor{MAGENTA}{cmyk}{0,1,0,0}
 \definecolor{YELLOW}{cmyk}{0,0,1,0}
}

\@ifundefined{showcaptionsetup}{}{%
 \PassOptionsToPackage{caption=false}{subfig}}
\usepackage{subfig}
\makeatother

\begin{document}

\preprint{Version 0.3 \today}

\title{Dynamical Tuning of Energy Transfer Efficiency on a Graphene Monolayer}

\author{Vasilios D. Karanikolas}

\affiliation{Semiconductor Photonics Group, School of Physics and CRANN,\\
 Trinity College Dublin, College Green 2, Dublin, Ireland}

\author{Cristian A. Marocico}

\affiliation{Semiconductor Photonics Group, School of Physics and CRANN,\\
 Trinity College Dublin, College Green 2, Dublin, Ireland}

\author{A. Louise Bradley}

\affiliation{Semiconductor Photonics Group, School of Physics and CRANN,\\
 Trinity College Dublin, College Green 2, Dublin, Ireland}

\date{\today}

\begin{abstract}
We present in this contribution a theoretical investigation of the
spontaneous emission and energy transfer rates between quantum systems
placed above a monolayer of conducting graphene. The conditions for
strong and weak coupling between a quantum system and the surface
plasmon-polariton of graphene are determined and, subsequently, we
focus exclusively on the weak coupling regime. We then calculate the
dispersion relation of the surface plasmon mode on graphene and, by
varying the chemical potential, show a good control of its resonance
frequency. Using a Green's tensor formalism, we calculate the spontaneous
emission and energy transfer rates of quantum systems placed near
the graphene monolayer. The spontaneous emission rate of a single
quantum system is enhanced by several orders of magnitude close to
the graphene monolayer and we show that this enhancement is due almost
exclusively to excitation of the surface plasmon mode. When considering
the energy transfer rate between two quantum systems, we find a similar
enhancement of several orders of magnitude close to the graphene monolayer.
The direct interaction between the donor and acceptor dominates when they
are close to each other, but is modified from its free-space behavior
due to the presence of the graphene monolayer. As the donor-acceptor
separation is increased, their direct interaction is overshadowed
by the interaction via the surface plasmon mode. Due to the large
propagation length of surface plasmon mode on graphene -- hundreds
of nanometers -- this enhancement of the energy transfer rate holds
over large donor-acceptor separations along the graphene monolayer.
\end{abstract}
\maketitle

\section{Introduction\label{sec:01}}

For the last two decades the field of plasmonics has grown intensively.
Confining light to sub-wavelength structures by exciting surface plasmon-polariton
(SPP) modes has various applications in biosensing devices, light
harvesting, optical nanoantennas and quantum information processing.
SPPs are collective oscillations of electrons and the electromagnetic
field that are excited at the interface between a dielectric and a
conductor and can propagate along that interface.\cite{Barnes2003}
In plasmonics, noble metals are routinely used as the conducting medium.
The main drawback of using noble metals in the applications mentioned
above is their large Ohmic losses.\cite{Dionne2005}

Graphene constitutes an alternative to using noble metals for plasmonic
applications.\cite{GarciadeAbajo2014,Grigorenko2012,Nikitin2014,Bludov2013,Low2014a}
It is a two-dimensional material possessing unique properties. This
atomically thick monolayer has superior electronic and mechanical
properties originating in part from its charge carriers of zero effective
mass that can travel for microns without scattering at room temperatures.\cite{castro2009}

An undoped graphene monolayer (GM) can absorb $\pi\alpha_{0}\thickapprox2.3$\%
of the light incident upon it, where $\alpha_{0}$ is the fine structure
constant.\cite{Nair2008} Patterned GM nanostructures can give rise
to $100$\% absorption at specific wavelengths, which can be tuned
through the applied voltage.\cite{Thongrattanasiri2012,Con2013,Nikitin2012}

Interactions between quantum emitters (QEs) and a GM have been investigated
intensively over the last few years, especially as regards the influence
of the GM on the spontaneous emission (SE) rate of a QE placed near
the GM.\cite{Hanson2013,Nikitin2011,Koppens2011,Tisler2013,Konstantatos2012}
It has been found that the SE rate can be enhanced by several order
of magnitude compared with its free space value, due to the excitation
of graphene plasmon (GP) modes. Furthermore, confinement of the GP
modes to one dimension (graphene ribbons)\cite{Koppens2011,Nikitin2012}
and zero dimensions (graphene nanodisks)\cite{Koppens2011,Manjavacas2012,Con2013}
can enhance the QE-GM interaction even more. In the case of zero-dimensional
confinement, the strong-coupling regime can be achieved between a
QE and graphene nanodisks.\cite{Koppens2011}

The presence of a second QE in the system will modify the emission
properties of the system, giving rise to super- and sub- radiant states.\cite{Gonzalez-Tudela2011,Dzsotjan1} Once more, the confinement
of the GP to one dimension (graphene ribbons)\cite{Huidobro2012}
or zero dimensions (graphene nanodisks)\cite{Manjavacas2012} further
enhances the interaction rates. The energy transfer (ET) rate between
a pair of QEs has also been investigated in the presence of a GM.\cite{Biehs2013} 

In this contribution we investigate the SE and ET rates of quantum
emitters placed near a graphene layer using a semianalytical Green's
tensor formalism.\cite{Dung2002} We use quantum emitters with optical
properties corresponding to real physical systems. The SE and ET rates
are competitive processes, thus an ET efficiency is introduced. This
quantity, which is a measure of the contribution of the energy transfer
rate to the total decay rate of the donor, shows that one can efficiently
transfer energy between QEs separated by distances of the order of
hundreds of nanometers, due to the excitation of GP modes on the GM.
The ET efficiency can be tuned, through gating of the GM, thus opening
opportunities for applications such as switching and sensing
devices,\cite{Schwierz2010,Mazzamuto2014,He2012} light harvesting,\cite{Lin2011} plasmonic rulers\cite{Wang2011a,Mazzamuto2014}
and quantum computing. \cite{Gullans2014} Furthermore we show that
the ET rate along the GM has two contributions, a Förster contribution,
and a GP contribution, both of which can be tuned. We also show that
the ET rate perpendicular to the GM is mainly characterized by the
penetration depth of the GP mode. 

The paper is structured as follows. We begin in Sec.~\ref{sec:02}
with a theoretical framework in which we introduce the optical properties
of the GM and the GP mode it supports (Sec.~\ref{sub:02.A}), and
we also investigate the different coupling regimes of the QE-GM system,
i.e.~strong or weak coupling (Sec.~\ref{sec:02.B}). In Sec.~\ref{sec:03}
we present the results of our calculations for the SE rate (Sec.~\ref{sub:03.A}),
the ET function (Sec.~\ref{sub:03.B}) and the ET efficiency (Sec.~\ref{sub:03.C}),
for different distance regimes and values of the chemical potential.
Finally, in Sec.~\ref{sec:04} we give a summary of the results and
the conclusions drawn. In addition, we include two Appendixes where
we present various expressions used in the main body of the paper.

\section{Theoretical Framework\label{sec:02}}

\begin{figure}
\includegraphics[width=0.48\textwidth]{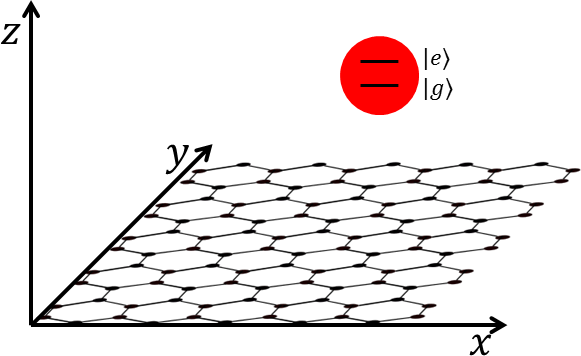}\caption{(Color online) A graphene monolayer in the $xy$-plane, with the quantum
emitter approximated as a two level system.\label{fig:01}}
\end{figure}

The model system considered in this paper is presented as a sketch
in Fig.~\ref{fig:01}.
 We consider an atomically thin monolayer of graphene in the $xy$-plane,
suspended in vacuum. Close to this graphene monolayer, we have either
a single quantum system, when investigating spontaneous emission,
or two quantum systems, for energy transfer investigations. We begin
our investigation by considering a conductivity model for graphene
and the GP properties.

\subsection{Graphene Conductivity and GP Properties\label{sub:02.A}}

\begin{figure}
\centering{}\includegraphics[width=0.48\textwidth]{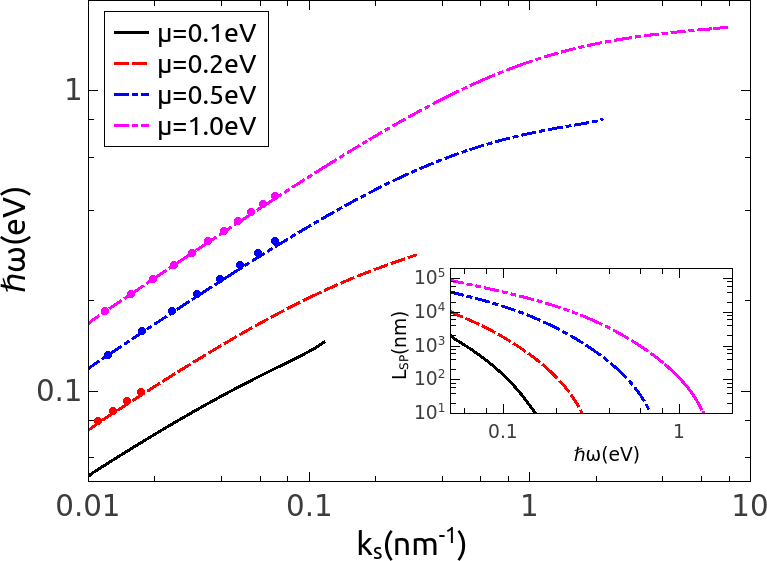}\caption{(Color online) GP dispersion relation in doped graphene, $\hbar\omega(k_{\text{s}}),$
for different values of the chemical potential $\mu$; the dots show
the quadratic approximation of the dispersion relation. The inset
shows the propagation length of the GP mode, $L_{\text{SP}}$, for
the same values of the chemical potential, $\mu$.\label{fig:02}}
\end{figure}

We calculate the graphene in-plane conductivity, $\sigma$, in the
random phase approximation (RPA).\cite{Jablan2009,Wunsch2006,Falkovsky2008}
This quantity is mainly controlled by electron-hole pair excitation
that can be divided into intraband and interband excitations. It can
be written as
\begin{equation}
\sigma=\sigma_{\text{intra}}+\sigma_{\text{inter}},\label{eq:01}
\end{equation}
where the intraband and interband contributions are,

\begin{equation}
\sigma_{\text{intra}}=\frac{2ie^{2}t}{\hbar\pi(\Omega+i\gamma)}\ln\bigg[2\cosh\bigg(\frac{1}{2t}\bigg)\bigg],\label{eq:02}
\end{equation}
\begin{equation}
\sigma_{\text{inter}}=\frac{e^{2}}{4\hbar}\bigg[\frac{1}{2}+\frac{1}{\pi}\arctan\bigg(\frac{\Omega-2}{2t}\bigg)-\frac{i}{2\pi}\ln\frac{(\Omega+2)^{2}}{(\Omega-2)^{2}+(2t)^{2}}\bigg].\label{eq:03}
\end{equation}
In the above we have introduced the dimensionless parameters $\Omega=\hbar\omega/\mu$,
$\gamma=E_{\text{S}}/\mu$ and $t=k_{\text{B}}T/\mu$. Here, $\mu$
is the chemical potential, $T$ is the temperature, and $E_{\text{S}}$
is the scattering energy. This scattering energy is related to the
relaxation time $\tau$ through $\tau=\hbar/E_{\text{S}}$. The $\sigma_{\text{intra}}$
term describes a Drude model response for intraband processes corrected
with a term $\gamma$, which accounts for impurities compromising
the electron's mobility. Throughout this paper we consider a temperature
$T=300\,\text{K}$ and a value of the relaxation time of $\tau=1\,\text{ps}$.\cite{Novoselov2004,Novoselov2005} Fig.~\ref{fig:02} shows the
dispersion relation, $\omega(k_{\text{s}})$, for different values
of the chemical potential, $\mu$. Due to the fact that retardation
effects dominate for wavevectors $k_{\text{s}}>k_{\text{F}}$ and
energies $\hbar\omega>2\mu$, the GP dispersion curves in Fig.~\ref{fig:02}
terminate at these values, where $k_{\text{F}}$ is the Fermi wavevector.
When the chemical potential, $\mu$, has the value $\mu=0$, the GM
absorbance has the value $\pi\alpha_{0}\thickapprox2.3$\%, where
$\alpha_{0}$ is the fine structure constant. This is the asymptotic
value of the doped GM for energies $\hbar\omega>2\mu$.
 Considering that the GM is surrounded by air, $\varepsilon_{1}=\varepsilon_{2}=1$
(free standing GM), the GP dispersion relation can be obtained form:
\begin{equation}
\frac{1}{\sqrt{k_{0}^{2}-k_{\text{SP}}^{2}}}=-\frac{2\pi\sigma}{\omega},\label{eq:04}
\end{equation}
where $k_{\text{SP}}$ is the GP wavevector.\cite{Koppens2011}
Because $k_{\text{SP}}\gg k_{0}$, we can simplify the dispersion
relation above, using only the $\sigma_{\text{intra}}$ contribution,
to obtain
\begin{equation}
k_{\text{SP}}=\frac{\hbar^{2}}{4e^{2}\mu k_{\text{B}}T\,\ln\left[2\cosh\left({\displaystyle \frac{\mu}{k_{\text{B}}T}}\right)\right]}\omega\left(\omega+\frac{i}{\tau}\right),\label{eq:05}
\end{equation}
which has as its main feature the quadratic dependence of the GP
wavevector on the frequency, when the intraband contributions dominate.
\cite{Jablan2009} This approximate quadratic dependence is shown
as dots in Fig.~\ref{fig:02}. Another feature of the GP dispersion
relation on graphene is the fact that the GP resonance frequency
is blue-shifted as the chemical potential increases.

The inset in Fig.~\ref{fig:02} shows the propagation length of the
surface plasmon along the graphene, $L_{\text{SP}}=1/\mbox{\text{Im}}(k_{\text{s}})$,
as a function of energy, for different values of the chemical potential,
$\mu$. It is evident from the plot that, depending on the value of
the chemical potential, $\mu$, this propagation length can reach
values as large as hundreds of microns at low frequencies. As the
energy is increased, however, the propagation length decreases rapidly,
because the GP then has sufficient energy to generate electron-hole
pairs and the dispersion relation is dominated by the interband contributions.\cite{GarciadeAbajo2014}

\subsection{Rabi splitting - strong coupling regime\label{sec:02.B}}

In order to ascertain whether the weak or strong coupling regime
applies for particular sets of parameters, we now consider a model
consisting of a single QE interacting with the GP mode of the GM.
To describe this system, we use a Jaynes-Cummings Hamiltonian\cite{Article}
of the form:
\begin{equation}
H=\frac{\hbar\omega}{2}\sigma_{z}+\hbar\omega_{\text{SP}}\,\hat{a}^{\dagger}\hat{a}+\hbar g(\hat{a}\sigma_{+}+\hat{a}^{\dagger}\sigma_{-}),\label{eq:06}
\end{equation}
where $\omega_{\text{SP}}$ is the GP frequency, $\hat{a}^{\dagger}$
and $\hat{a}$ are the creation and annihilation operators of the
plasmon mode, $\omega$ is the transition frequency of the QE between
its ground and excited state, $\sigma_{+}$ and $\sigma_{-}$ are
the raising and lowering operators of the QE, and $g$ is the coupling
constant between the QE and the GP mode of the GM. The coupling constant
$g$ is given by\cite{Hummer2013} 
\begin{align}
\left|g(\omega)\right|^{2}= & \frac{1}{\hbar\pi\varepsilon_{0}}\frac{\omega^{2}}{c^{2}}\mathbf{\hat{d}}^{T}\text{Im}\mathfrak{G}_{\text{SP}}(\omega,\mathbf{r}_{\text{QE}},\mathbf{r}_{\text{QE}})\mathbf{\hat{d}}\nonumber \\
= & \gamma_{0}\frac{3c}{\omega}\mathbf{\hat{d}}^{T}\text{Im}\mathfrak{G}_{\text{SP}}(\omega,\mathbf{r}_{\text{QE}},\mathbf{r}_{\text{QE}})\mathbf{\hat{d}},\label{eq:07}
\end{align}
where $\mathfrak{G}_{\text{SP}}\left(\omega,\mathbf{r}_{\text{QE}},\mathbf{r}_{\text{QE}}\right)$
is the GP part of the Green's tensor, derived in Appendix~\ref{app:A},
Eq.~(\ref{eq:A02a}), $\gamma_{0}$ is the Einstein A-coefficient,
$\gamma_{0}=\omega^{3}d^{2}/(3\pi\hbar\varepsilon_{0}c^{3})$ and
$\mathbf{d}$ is the transition dipole moment of the QE positioned
at $\mathbf{r}_{\text{QE}}=(0,0,z)$. In this section, we consider
$\gamma_{0}=5\times10^{-8}\,\text{fs}^{-1}$ and we consider the transition
dipole moment, $\mathbf{d}$, of the QE to be oriented along $z$.
Since we are interested in the coupling between the GP mode and the
QE we calculate the GP contribution to Eq.~(\ref{eq:A02a}), by
extracting the pole contribution and we obtain 
\begin{equation}
\mathfrak{G}_{\text{S},zz}^{(11)\text{SP}}(z,\omega)=-\frac{1}{4}\frac{\big(1-1/\alpha^{2}\big)}{\alpha k_{0}}\text{e}^{-2iz/\alpha},\label{eq:08}
\end{equation}
where $\alpha=2\pi\sigma/c$. 

The Hamiltonian from Eq.~(\ref{eq:06}) couples the states $\left|e\right\rangle \otimes\left|0\right\rangle $
and $\left|g\right\rangle \otimes\left|1\right\rangle $ to the dressed
states $\left|1\right\rangle $ and $\left|2\right\rangle $ with
energies,\begin{subequations}\label{eqs:09} 
\begin{equation}
E_{1}=\frac{\hbar\omega_{\text{SP}}}{2}-\frac{\hbar}{2}\sqrt{\delta^{2}+4g^{2}},\label{eq:09a}
\end{equation}
\begin{equation}
E_{2}=\frac{\hbar\omega_{\text{SP}}}{2}+\frac{\hbar}{2}\sqrt{\delta^{2}+4g^{2}},\label{eq:09b}
\end{equation}
\end{subequations}where $\delta=\omega_{\text{SP}}-\omega$, is the
detuning between the GP mode resonant frequency and the transition
frequency of the QE. The energy states are separated by $\Omega=\sqrt{\delta^{2}+4g^{2}}$,
which gives the value of the Rabi splitting. As an example, if we
consider the case where the QE is positioned at $z=10\,\text{nm}$
above a GM with a chemical potential equal to the transition energy
of the QE, $\mu=\hbar\omega=0.5\,\text{eV}$, the Rabi splitting,
at $\delta=0$, has a value $2g(\omega)=0.12\,\text{eV}$.

In order to further investigate the weak and strong coupling regimes,
we will analyze the dependence of the coupling constant, $g$, on
the various parameters involved, namely the value of the chemical
potential, $\mu$, the emission frequency of the QE, $\omega$, and
the distance of the QE to the GM, $z$. Considering $\omega=\omega_{\text{SP}}$,
i.e.~zero detuning, the criterion for having strong coupling is whether
or not the absorption spectrum of the system exhibits two peaks of
different frequencies.\cite{Andreani1999,Hummer2013} This condition
is fulfilled if
\begin{equation}
\left|g\right|>\frac{1}{4}\left|\gamma_{\text{LSW}}-\kappa\right|,\label{eq:10}
\end{equation}
where $\gamma_{\text{LSW}}$ represents the \emph{lossy surface waves}
(LSW) contribution, which are non-propagating evanescent modes relaxing
through Ohmic losses, and $\kappa$ is the width of the $g(\omega)^{2}$
spectrum.

\begin{figure*}
\subfloat[\label{fig:03a}]{\includegraphics[width=0.48\textwidth]{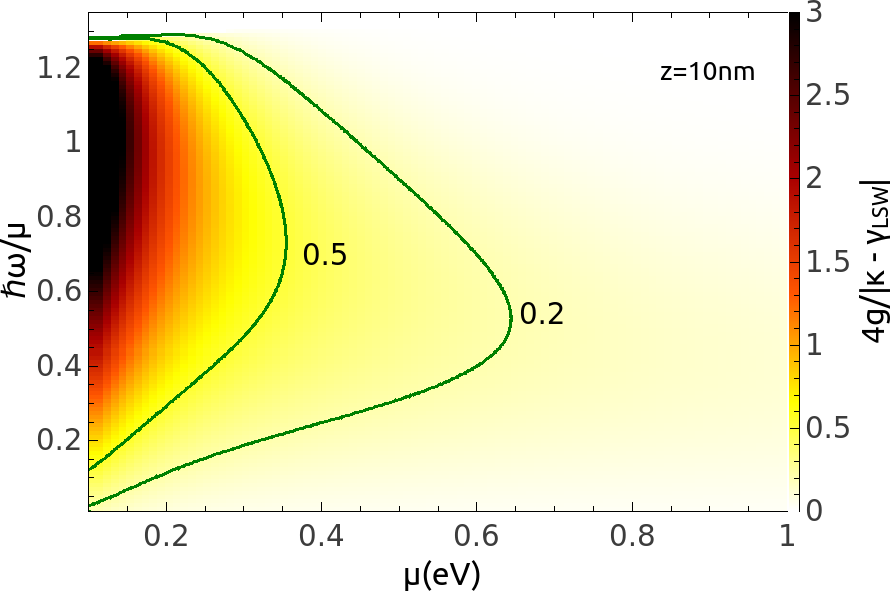}}\hspace*{\fill}\subfloat[\label{fig:03b}]{\includegraphics[width=0.48\textwidth]{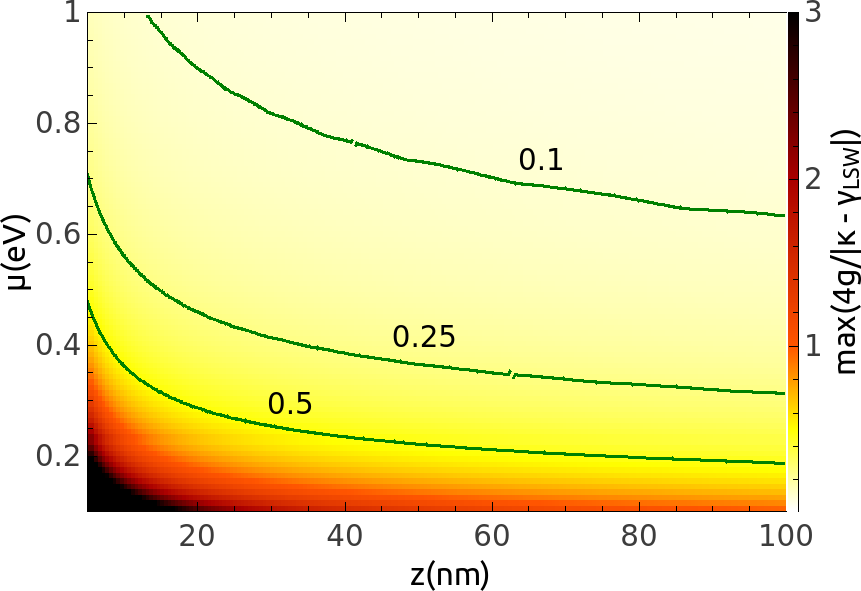}}\caption{(Color online) Contour plots of the quantity $\mathfrak{D}=4\left|g\right|/\left|\gamma_{LSW}-\kappa\right|$
as a function of (a) chemical potential, $\mu$, and emission energy,
$\hbar\omega$ normalized by the chemical potential, for fixed QE position $z=10\,\text{nm}$, (b) chemical
potential, $\mu$, and QE position, $z$; each point reflects the
maximum value of $\mathfrak{D}(\omega)$.\label{fig:03}}
\end{figure*}

In Fig.~\ref{fig:03a} we present a contour plot of the quantity
$\mathfrak{D}=4\left|g\right|/\left|\gamma_{LSW}-\kappa\right|$ as
a function of the chemical potential, $\mu$, and the emission energy,
$\hbar\omega$, for a fixed position of the QE, $z=10\,\text{nm}$.
 Although from condition (\ref{eq:10}) when $\mathfrak{D}<1$ we
are in the weak coupling regime, this condition might not be sufficient
under some experimental conditions, \cite{Article} and we thus consider
the more stringent inequality $\mathfrak{D}\lesssim0.5$ as giving
the weak coupling condition.  As is evident from Fig.~\ref{fig:03a},
for chemical potential values $\mu<0.3\,\text{eV}$, there exists
a frequency region where we have $\mathfrak{D}\geq$0.5 and the weak
coupling approximation needed to calculate the SE and ET rates is
no longer valid. This region where the strong coupling dominates corresponds
to THz frequencies, a range outside the scope of our investigation.
For chemical potential values $\mu>0.6\,\text{eV}$, on the other
hand, the quantity $\mathfrak{D}$ has values below 0.2, at all frequencies, well within
the weak-coupling regime.

Fig.~\ref{fig:03b} examines the maximum value of $\mathfrak{D}$
for different values of the chemical potential, $\mu$, and at different
positions of the QE above the QM. For each point, we calculate the
maximum value of $\mathfrak{D}$ as a function of the emission energy
of the QE, $\hbar\omega$. This represents, therefore, the worst-case
scenario for weak coupling, since at all other frequencies, $\mathfrak{D}$
will be smaller than the values depicted in Fig.~\ref{fig:03b}. It can be seen that the SC regime is only accessed for values of $\mu<0.4\,\text{eV}$ at certain frequencies. Throughout the rest of the paper   we only consider frequencies ranges which remain outside the SC regime for all values of $\mu$ and $g$ explored.

\section{Results and Discussion\label{sec:03}}

\subsection{Spontaneous emission rate\label{sub:03.A}}

The decay rate $\gamma$ is proportional to the strength of the coupling
between the transition dipole matrix element and the electromagnetic
modes acting on it. The normalized SE has the expression\cite{Dung2002}
\begin{equation}
\tilde{\gamma}=\frac{\gamma}{\gamma_{0}}=n_{i}+\frac{6\pi c}{\omega}\textrm{Im}\left[\mathbf{n}_{\text{QE}}\cdot\,\mathfrak{G}_{\text{s}}(\mathbf{r},\mathbf{r},\omega)\cdot\mathbf{n}_{\text{QE}}\right],\label{eq:11}
\end{equation}
where $\gamma_{0}$ is Einstein's $A$-coefficient, $\mathbf{n}_{\text{QE}}$ is
a unit vector along the direction of the transition dipole moment
of the emitter, and $\mathfrak{G}\left(\mathbf{r},\mathbf{r},\omega\right)$ is
given by (\ref{eqs:A02}).

\begin{figure*}
\subfloat[\label{fig:04a}]{\centering{}\includegraphics[height=0.25\textheight]{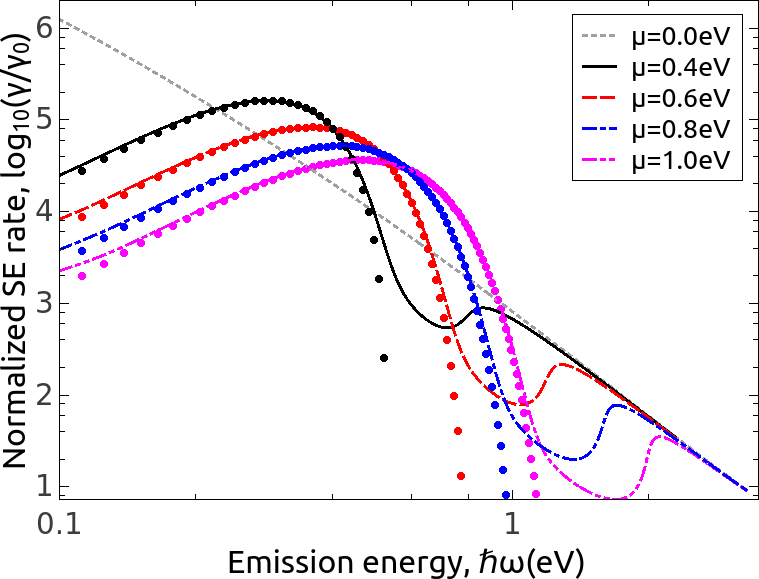}}\hspace*{\fill}\subfloat[\label{fig:04b}]{\centering{}\includegraphics[height=0.25\textheight]{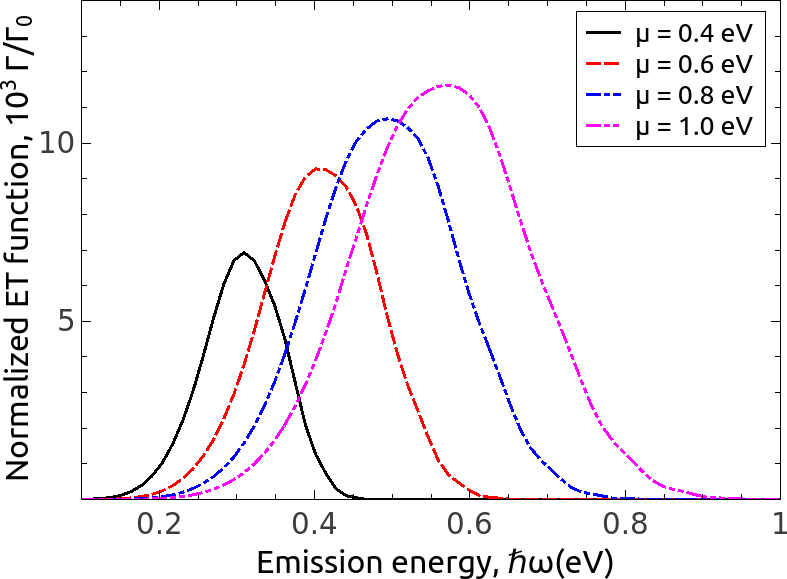}}\caption{(Color online) (a) SE rate of a QE, positioned at $\mathbf{r}=(0,0,10\,\text{nm})$,
as a function of its emission frequency for different values of the
chemical potential, $\mu$. (b) Normalized ET function, $\tilde{\Gamma}(\mathbf{r}_{\text{A}},\mathbf{r}_{\text{D}},\omega)$,
between a donor and acceptor QE as a function of frequency, for fixed
donor and acceptor positions, $\mathbf{r}_{\text{D}}=(0,0,10\,\text{nm})$
and $\mathbf{r}_{\text{A}}=(100\,\text{nm},0,10\,\text{nm})$ respectively,
and different values of the chemical potential, $\mu$.\label{fig:04}}
\end{figure*}

In Fig.~\ref{fig:04a} we have plotted the normalized SE rate, $\tilde{\gamma}$,
of a QE at a fixed position, $\mathbf{r}=(0,0,10\,\text{nm})$ above
the GM, as a function of the QE's emission energy, $\hbar\omega$,
for different values of the chemical potential, $\mu$.
 In general the SE rate has a peak at an energy below $\mu$. As the
energy is further increased, the SE rate drops dramatically before
finally recovering to a value independent of $\mu$, when the energy
is above $2\mu$. As we increase the value of $\mu$, the GP peak
blue-shifts and is broadened, and its value decreases. The general
drop in the SE rate is most visible starting with values of the chemical
potential of $\mu>0.4\,\text{eV}$, and it occurs between the energies
$\hbar\omega=\mu$ and $\hbar\omega=2\mu$. This drop is due to interband
transitions when the QE relaxes through lossy channels. At emission
energies $\hbar\omega>2\mu$ the emission is determined by interband
contributions and GP excitations become unimportant, as the dispersion
relations in Fig.~\ref{fig:02} show. At these energies the SE rate
follows the same behavior as for the case of undoped graphene, $\mu=0$
eV, as seen in Fig.~\ref{fig:04a}. Moreover, we can see that the
main contribution to the peak in the normalized SE rate, $\tilde{\gamma}$,
comes from the GP contribution, which is denoted by the circular
symbols in Fig. \ref{fig:04a}. The maximum value of $\mathfrak{D}$
is 0.41 at $\mu=0.4\,\text{eV}$, thus placing us within the weak
coupling regime.

\subsection{Energy transfer function\label{sub:03.B}}

In this section we investigate the influence of the GM on the energy
transfer process between a pair of QEs, a donor and an acceptor. The
normalized energy transfer function which we investigate in this section
is given as
\begin{equation}
\tilde{\Gamma}=\frac{\Gamma}{\Gamma_{0}}=\frac{\left|\mathbf{n}_{\text{A}}\cdot\mathfrak{G}\left(\mathbf{r}_{\text{A}},\mathbf{r}_{\text{D}},\omega\right)\cdot\mathbf{n}_{\text{D}}\right|^{2}}{\left|\mathbf{n}_{\text{A}}\cdot\mathfrak{G}_{0}\left(\mathbf{r}_{\text{A}},\mathbf{r}_{\text{D}},\omega\right)\cdot\mathbf{n}_{\text{D}}\right|^{2}},\label{eq:12}
\end{equation}
(see also (\ref{eq:B05})).

Fig.~\ref{fig:04b} shows the normalized energy transfer function,
$\tilde{\Gamma}$, as a function of frequency for different values
of the chemical potential, $\mu$, and when both the donor and acceptor
transition dipole moments are oriented perpendicular to the GM, i.e.~$zz$-orientation.
The donor and acceptor positions are fixed at $\mathbf{r}_{\text{D}}=(0,0,10\,\text{nm})$
and $\mathbf{r}_{\text{A}}=(100\,\text{nm},0,10\,\text{nm})$, respectively.
As for the case of the SE rate in \ref{fig:04a}, the normalized ET
function, $\tilde{\Gamma}$, is enhanced close to the GP frequency
and in general for frequencies $\hbar\omega<\mu$, where the intraband
transitions dominate. For frequencies $\hbar\omega>\mu$ the energy
transfer rate decreases due to the losses generated by the interband
contributions.

\begin{figure*}
\subfloat[\label{fig:05a}]{\includegraphics[height=0.25\textheight]{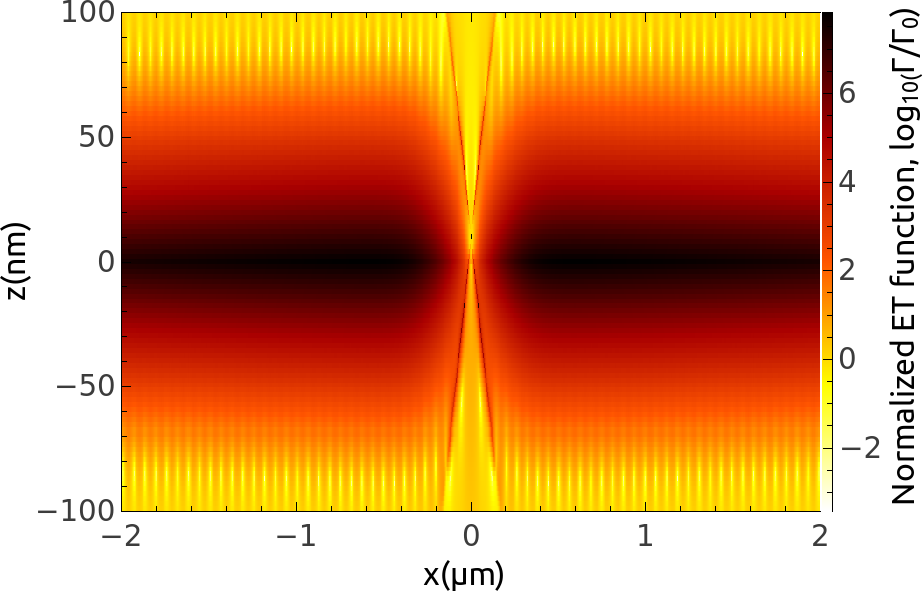}}\hspace*{\fill}\subfloat[\label{fig:05b}]{\includegraphics[height=0.25\textheight]{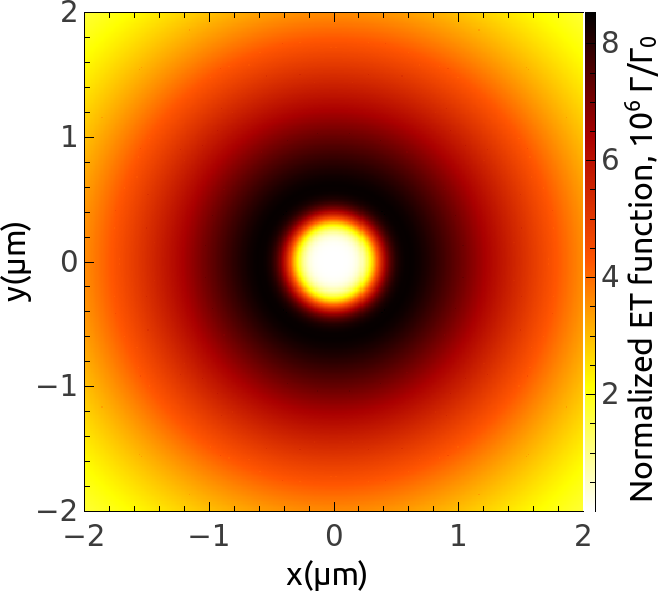}}\caption{(Color online) Contour plots of the normalized ET function, in the
(a) $xz$-plane and (b) $xy$-plane when $z=10\,\text{nm}$, for a
donor positioned at $\mathbf{r}_{\text{D}}=(0,0,10\,\text{nm})$ above
a GM. The dielectric permitivitty of the surrounding media is $\varepsilon_{1}=\varepsilon_{2}=1$.
The chemical potential of the graphene monolayer is $\mu=1.0\,\text{eV}$.
The emission frequency of the donor is $\omega=0.8\,\text{fs}^{-1}$
($\lambda=2.3\,\text{\ensuremath{\mu}m}$). Both donor and accceptor
have their transition dipole moments oriented along the $z$-axis.\label{fig:05}}
\end{figure*}

Fig.~\ref{fig:05} presents contour plots of the normalized ET function,
$\tilde{\Gamma}$, as a function of the position of the acceptor in
(a) the $xz$-plane and (b) the $xy$-plane 10~nm above the GM, when
the donor position is fixed at $\mathbf{r}_{\text{D}}=(0,0,10\,\text{nm})$,
the transition energy is $\hbar\omega=0.52\,\text{eV}$ ($\lambda=2.3\,\text{\ensuremath{\mu}m}$)
and the chemical potential is $\mu=1.0\,\text{eV}$.
 In Fig.~\ref{fig:05a} the normalized ET function has large values
when the acceptor is close to the GM and decreases as the acceptor
distance is increased. This behavior is due to the fact that the field
is highly confined in the $z$-direction at the surface of the GM,
with a penetration depth of $\delta_{\text{SP}}=10\,\text{nm}$, or
$\delta_{\text{SP}}/\lambda=4\cdot10^{-3}$. The fringes visible in
Fig.~\ref{fig:05a} are due to the constructive and destructive interference
between the direct and scattering terms in the Green's tensor, cf
Eq.~(\ref{eqs:A01}). This effect is more profound due to the dipole
moment orientations of the QEs, along the $z$-axis. Fig.~\ref{fig:05b}
shows that the normalized ET function has cylindrical symmetry in
the $xy$-plane, due to the orientation of both donor and acceptor
transition dipole moments along the $z$-axis. Furthermore, we see
that the normalized ET function has a peak value at a distance of
about $400$~nm, which is the propagation length of the GP mode for
the particular set of parameters used in this calculation.

\begin{figure*}
\subfloat[\label{fig:06a}]{\includegraphics[width=0.48\textwidth]{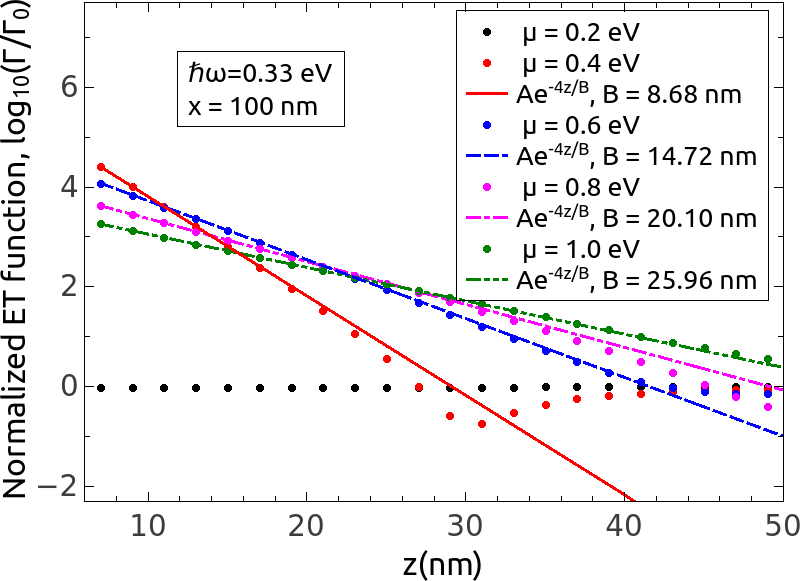}}\hspace*{\fill}\subfloat[\label{fig:06b}]{\includegraphics[width=0.48\textwidth]{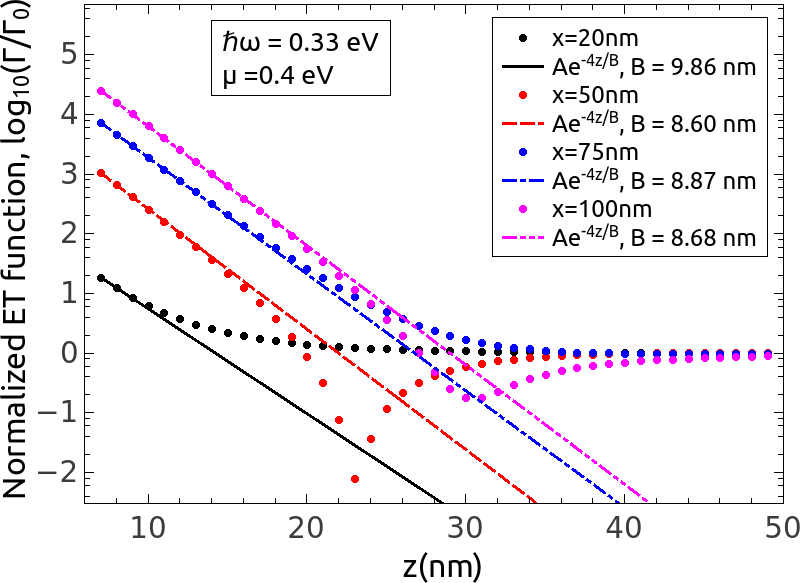}}\caption{(Color online) Normalized ET function, $\tilde{\Gamma}(\mathbf{r,s},\omega)$,
as a function of the donor and acceptor distance to the graphene monolayer,
for a fixed transition energy, $\hbar\omega=0.33\,\text{eV}$. (a)
For various values of the chemical potential, $\mu$, when the donor
position is $\mathbf{r}_{D}=(0,0,z)$ and the acceptor position is
$\mathbf{r}_{A}=(100\,\text{nm},0,z)$; (b) For various in-plane distances,
$x$, when the donor position is $\mathbf{r}_{D}=(0,0,z)$ and the
acceptor position is $\mathbf{r}_{A}=(x,0,z)$, for a fixed value
of the chemical potential, $\mu=0.4\,\text{eV}$. The symbols represent
the full simulation data, while the lines are fits with the exponential
function in the legend. Both donor and acceptor have their transition
dipole moments oriented along the $z$-axis.\label{fig:06}}
\end{figure*}

In Fig.~\ref{fig:06a} we present the $z$-dependence of the normalized
ET function, $\tilde{\Gamma}$, for a donor located at $\mathbf{r}_{\text{D}}=(0,0,z)$
and an acceptor at $\mathbf{r}_{\text{A}}=(100\,\text{nm},0,z)$,
for various values of the chemical potential, and a fixed transition
energy, $\hbar\omega=0.33\,\text{eV}$.
 As we have already pointed out, for different values of the chemical
potential, $\mu$, the position of the maximum in the normalized ET
function blueshifts as the value of the chemical potential is increased,
cf. Fig.~\ref{fig:04b}. Thus, for the smallest value of the chemical
potential, $\mu=0.2\,\text{eV}$, the enhancement of the ET function
is negligible to non-existent. As the value of the chemical potential
is increased, the ET function has values a few orders of magnitude
larger than in free-space when the donor-acceptor pair is close to
the GM. As the donor-acceptor pair is moved further from the GM --
increasing $z$ -- the ET function drops off exponentially, as the
figure shows. The continuous lines in Fig.~\ref{fig:06a} represent
fits of the calculated ET function (represented by symbols) with the
exponential function $Ae^{-4z/B}$. The factor $4$ in this expression
has two sources: (i) the $z$ distance of both donor and acceptor
to the GM is varied and (ii) the ET function depends on the square
of the field, see Eq.~(\ref{eq:12}); each of these gives a doubling
of the exponent, for a total of $2\times2=4$. Furthermore, because
the donor-acceptor separation is kept constant at $x=100\,\text{nm}$,
the free-space ET function is also constant and, hence, does not influence
the exponential behavior. Table~\ref{tab:01} shows the values of
the parameter $B_{\text{ET}}$ extracted from the fit, together with
the analogous parameter $B_{\text{SE}}$ extracted from fitting the
SE dependence (the numerical factor in the exponent is 2 in this case,
rather than 4, due to the fact that the SE rate depends linearly on
the electric field; data not shown) and the penetration depth of the
GP in the air above the graphene, calculated as $\delta_{\text{SP}}=1/\text{Im}\left(k_{z}^{\text{SP}}\right)$.
\begin{table}
\centering{}\caption{\label{tab:01}}
\begin{tabular}{|c|c|c|c|c|}
\hline 
$\hbar\omega$ (eV) & $\mu$ (eV) & $\delta_{\text{SP}}$ (nm) & $B_{\text{SE}}$ (nm) & $B_{\text{ET}}$ (nm)\tabularnewline
\hline 
\hline 
0.33 & 0.4 & 8.2 & 8.50 & 8.68\tabularnewline
\hline 
0.33 & 0.6 & 14.06 & 14.38 & 14.72\tabularnewline
\hline 
0.33 & 0.8 & 19.48 & 20.04 & 20.10\tabularnewline
\hline 
0.33 & 1.0 & 24.76 & 25.47 & 25.96\tabularnewline
\hline 
\end{tabular}
\end{table}
 As this table shows, the values of the parameters $B_{\text{SE}}$
and $B_{\text{ET}}$ match closely the calculated penetration depths
of the GPs. This suggests that the main channel for enhancing the
ET function between the QEs is the GP resonance. As the separation
between the two QEs becomes smaller and smaller, this relaxation channel
becomes less and less important and the direct interaction dominates
at distances below $x=20\,\text{nm}$, as panel \ref{fig:06b} shows.
In this panel we consider the influence of the in-plane distance between
the donor and acceptor, for a fixed value of the chemical potential,
$\mu=0.4\,\text{eV}$ and fixed transition energy, $\hbar\omega=0.33\,\text{eV}$.
We again fit the calculated ET function data (symbols) with the expression
$Ae^{-4z/B}$. As one can see, above $x=50\,\text{nm}$ the $B_{\text{ET}}$
parameter extracted is very close to the penetration depth of the
GP, $\delta_{\text{SP}}$. When the donor and acceptor in-plane distance
is smaller than $x=50\,\text{nm}$, the influence of the GP mode on
the ET function is less profound, and, as the donor-acceptor pair
is moved away from the GM, it recovers its free-space interaction
at shorter distances. This is due to the fact that the influence of
the homogeneous part of the Green's tensor dominates over the scattered
part, cf Eq.~\ref{eqs:A01}; this effect will be further discussed
in Sec.~\ref{sub:03.C}. In this section the quantity $\mathfrak{D}$
has the largest value of 0.48 for $\mu=0.4\,\text{eV}$, $\hbar\omega=0.33\,\text{eV}$
at a donor-GM distance of 8~nm, confirming that the system is in the weak coupling regime.

\subsection{Energy transfer efficiency\label{sub:03.C}}

In the last section we investigated the SE and ET functions. When
the donor dipole is excited it has two ways of relaxing to the ground
state: it can either transfer its excitation energy to the acceptor
dipole with an ET rate $k_{\text{ET}}$, or it can relax with decay
rate $k_{\text{SE}}$, where it is assumed that there is no non-radiative
decay, i.e.~the intrinsic quantum yield of the donor is $Y_{0}=1$.
The decay rate $k_{\text{SE }}$ takes account of photon emission
into the far-field, coupling to GP modes and losses in the GM. The
SE and ET processes are, therefore, in competition with each other
and we introduce an energy transfer efficiency to describe this competition.

We define the energy transfer efficiency $\eta$ as 
\begin{equation}
\eta=\frac{k_{\text{ET}}}{k_{\text{SE}}+k_{\text{ET}}}.\label{eq:13}
\end{equation}
This quantity gives the relative contribution of the energy transfer
process to the total decay rate of the donor. If the ET efficiency,
$\eta$, has a value $\eta>50\%$, then the decay of the excited state
of the donor occurs mainly by energy transfer to the acceptor, rather
than relaxation into photon or GP modes.

We next consider a donor-acceptor pair. The donor emission and acceptor
absorption spectra are both given by a Gaussian distribution 
\begin{equation}
A_{\text{q}}\,\text{e}^{-(\lambda-\lambda_{\text{q}})^{2}/\Delta\lambda_{\text{q}}^{2}},\label{eq:14}
\end{equation}
where $\text{q}=D$ represents the donor and $\text{q}=A$ represents
the acceptor. $A_{\text{q}}$ is a normalization constant, $\lambda_{\text{q}}$
gives the position of the spectral peak and $\Delta\lambda_{\text{q}}$
is related to the full width at half maximum (FWHM) of the spectrum.
The donor emission peak and acceptor absorption maximum coincide at
$\lambda_{\text{D}}=\lambda_{\text{A}}=2\;\text{\ensuremath{\mu}m}$.
There are a variety of emitters at this wavelength, such as quantum
dots and synthesized molecules.\cite{Pietryga2004,Treadway2001a}
The normalization constant of the donor emission spectrum, $f_{\text{D}}(\lambda)$,
is given as $A_{\text{D}}^{-1}=\int_{0}^{\infty}\text{d}\lambda f_{\text{D}}(\lambda)$.
The width of the spectrum will be $\Delta\lambda_{\text{D}}=20\;\text{nm}$,
a reasonable value for a typical QE. The normalization constant for
the acceptor absorption spectrum is $A_{\text{A}}=0.021\;\text{nm}^{2}$,
while the width is $\Delta\lambda_{\text{A}i}=50\;\text{nm}$.

The Förster radius, $R_{0}$, is defined as the donor-acceptor separation
at which the energy transfer efficiency $\eta$ is $50\%$. The Förster
radius can be calculated from the spectral overlap and has a value
of $19\,\text{nm}$ in free-space. $R_{0}$ is calculated from the
spectral overlap of the normalized donor emission, $f_{\text{D}}$,
and acceptor absorption, $\sigma_{\text{A}}$, spectra as
\begin{equation}
R_{0}=\left[\frac{3c}{32\pi^{4}n_r^{4}}\int\limits _{0}^{\infty}\text{d}\lambda\lambda^{2}f_{\text{D}}\left(\lambda\right)\sigma_{\text{A}}\left(\lambda\right)\right]^{1/6}\label{eq:15}
\end{equation}
where $n_r$ is the refractive index of the host medium, in our case
air with $n_r=1$.

\begin{figure*}
\subfloat[\label{fig:07a}]{\includegraphics[width=0.48\textwidth]{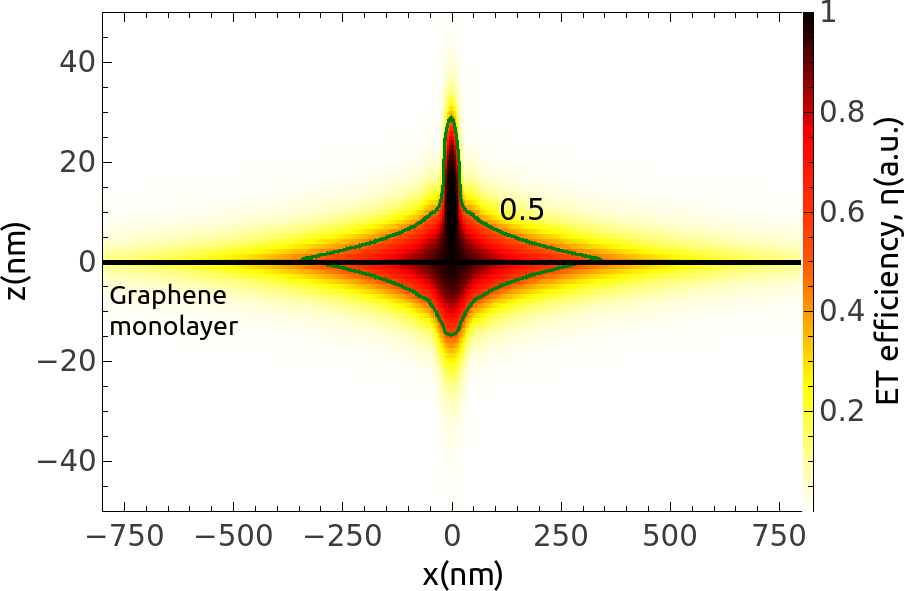}}\hfill{}\subfloat[\label{fig:07b}]{\includegraphics[width=0.48\textwidth]{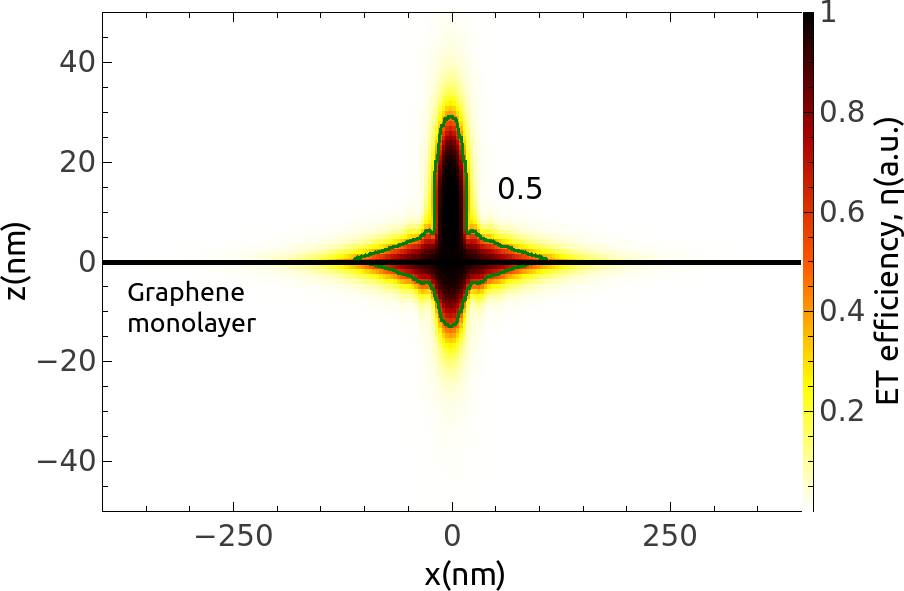}}

\caption{(Color online) Contour plot of the ET efficiency, $\eta$, between
a pair of QEs, as a function of acceptor position for a fixed position
of the donor at $\mathbf{r}_{\text{D}}=(0,0,10\,\text{nm})$ and different
values of the chemical potential, (a) $\mu=1\,\text{eV}$ and (b)
$\mu=0.6\,\text{eV}$. The green line gives the $\eta=\text{50\%}$
values.\label{fig:07}}
\end{figure*}

In Fig.~\ref{fig:07} we present contour plots of the ET efficiency
for the donor-acceptor pair, with spectral properties described above;
the donor and acceptor positions are fixed at $\mathbf{r}_{\text{D}}=(0,0,10\,\text{nm})$,
and $\mathbf{r}_{\text{A}}=(x,0,z)$, respectively.
 The chemical potential takes on two values, (\ref{fig:07a}) $\mu=1.0\,\text{eV}$
and (\ref{fig:07b}) $\mu=0.6\,\text{eV}$. When $\mu=1.0\,\text{eV}$
the emission and absorption spectrum overlap strongly with the ET
function. For this case the ET efficiency, $\eta$, has values above
$70\%$ even for separations along the GM as large as $100\,\text{nm}$,
and the 50\% efficiency distance is around $300\,\text{nm}$. This
value is very large compared to the free-space Förster radius of $R_{0}=19\,\text{nm}$.
When the value of the chemical potential is $\mu=0.6\,\text{eV}$,
the ET efficiency, $\eta$, has values above $50\,\%$ for separations
above $100\,\text{nm}$ but now the overlap between the emission and
absorption spectra and the ET function is poorer, thus showing a diminished
effect. The large confinement of light at the atomically thin GM surface
can be used to efficiently transfer energy between a pair of QEs over
large separations. Furthermore, this ET efficiency, $\eta$, can be
controlled through gating of the GM, thus opening striking opportunities
for possible applications, such as switching and sensing devices,\cite{Schwierz2010,Mazzamuto2014,He2012}
light harvesting,\cite{Lin2011} plasmonic rulers\cite{Wang2011a,Mazzamuto2014}
and quantum computing.\cite{Gullans2014}

\begin{figure*}
\subfloat[\label{fig:08a}]{\includegraphics[width=0.48\textwidth]{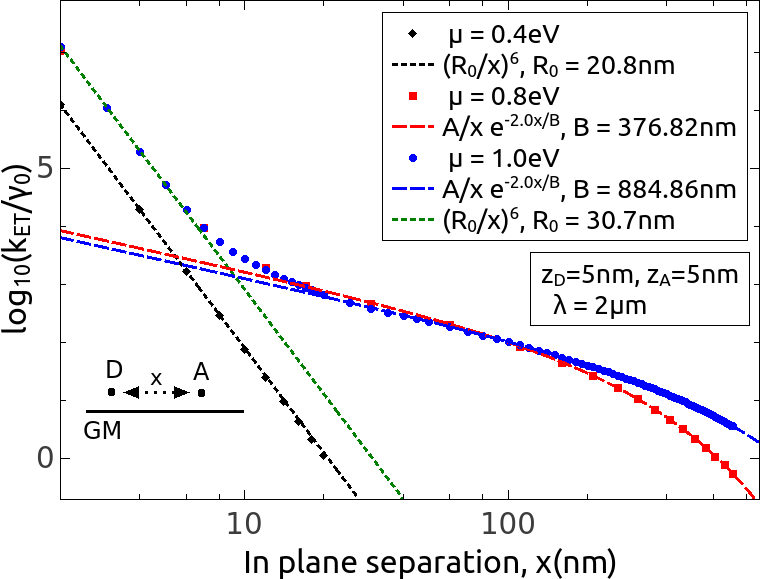}}\hfill{}\subfloat[\label{fig:08b}]{\includegraphics[width=0.48\textwidth]{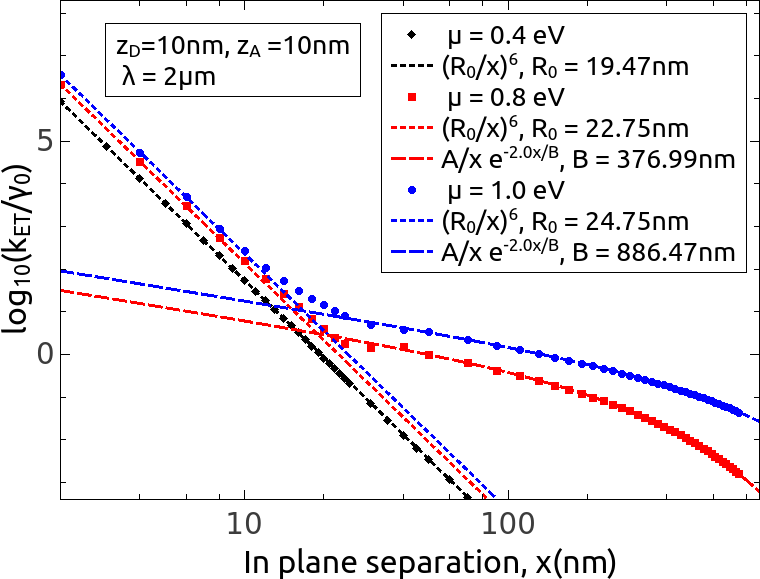}

}

\subfloat[\label{fig:08c}]{\includegraphics[width=0.48\textwidth]{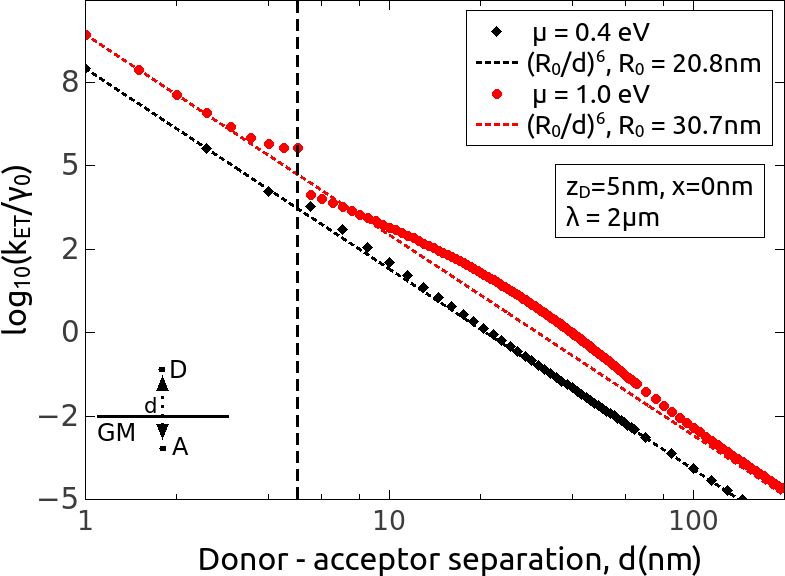}}\hfill{}\subfloat[\label{fig:08d}]{\includegraphics[width=0.48\textwidth]{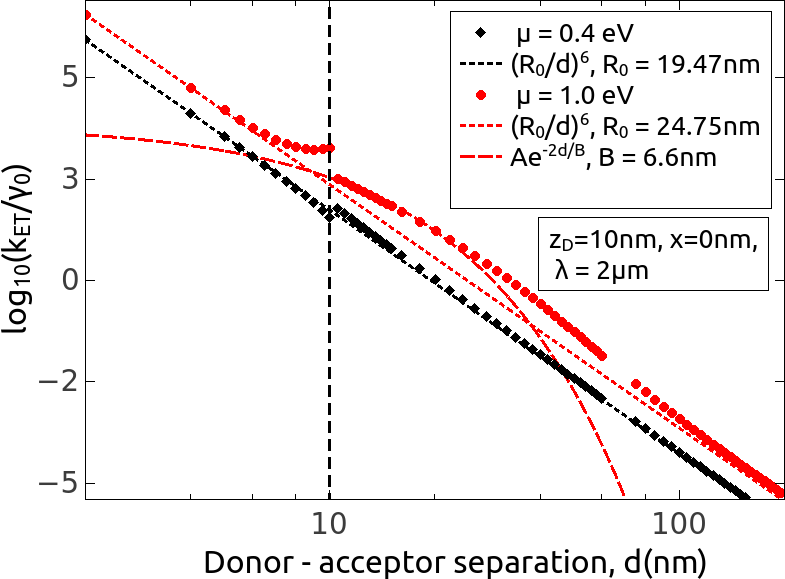}

}\caption{(Color online) (a-b) ET rate, $k_{\text{ET}}$, as a function of the
donor-acceptor in-plane separation, $x$, for fixed donor positions,
(a) $\mathbf{r}_{\text{D}}=(0,0,5\,\text{nm})$ and (b) $\mathbf{r}_{\text{D}}=(0,0,10\,\text{nm})$
respectively, and variable acceptor position, $\mathbf{r}_{\text{A}}=(x,0,5\,\text{nm})$
and $\mathbf{r}_{\text{A}}=(x,0,10\,\text{nm})$ respectively, for
different values of the chemical potential, $\mu$. (c-d) ET rate,
$k_{\text{ET}}$, as a function of the donor-acceptor separation,
$d$, for fixed donor positions, (c) $\mathbf{r}_{\text{D}}=(0,0,5\,\text{nm})$
and (d) $\mathbf{r}_{\text{D}}=(0,0,10\,\text{nm})$ respectively,
and variable acceptor position $\mathbf{r}_{\text{A}}=(0,0,z_{A})$,
with $d=z_{\text{D}}-z_{\text{A}}$ and for different values of the
chemical potential, $\mu$.\label{fig:08}}
\end{figure*}

We next consider the behavior of the ET rate, $k_{\text{ET}}$, as
a function of the in-plane separation between donor and acceptor.
Figs.~\ref{fig:08a} and \ref{fig:08b} show the ET rate, $k_{\text{ET}}$,
as a function of the in-plane separetion between the donor and acceptor,
when their elevation above the GM is (a) $z_\text{D}=z_\text{A}=5\,\text{nm}$ and (b)
$z_\text{D}=z_\text{A}=10\,\text{nm}$ on the same side of the GM.
 For both panels, we have fitted the near field with a Förster-type
model, $(R_{0}/x)^{n}$, where $R_{0}$ is the Förster radius. At
small separations the fit yields the values $n=6$ typical of Förster ET, but we see
that the Förster radius is modified from the free-space value.
The fact that at small separations, $x<10\,\text{nm}$, for $z_{\text{D}}=z_{\text{A}}=5\,\text{nm}$
and $x<20\,\text{nm}$, for $z_{\text{D}}=z_{\text{A}}=10\,\text{nm}$,
the ET rate, $k_{\text{ET}}$, follows an $n=6$ dependence shows
that the homogeneous part of the Green's tensor dominates, cf. (\ref{eq:A01a}),
modified by the donor-acceptor interaction with the GM. Thus, there
is an enhancement of the Förster radius, which depends on the donor-acceptor
distances from the GM and the value of the chemical potential, $\mu$.
In Fig.~\ref{fig:08a} the Förster radius has a value of $R_{0}=30.7\,\text{nm}$
for chemical potential values $\mu=0.8\,\text{eV}$ and $\mu=1.0\,\text{eV}$.
As the chemical potential value drops, the Förster radius decreases
to $R_{0}=20.8\,\text{nm}$ for $\mu=0.4\,\text{eV}$, approaching
the free-space value of the Förster radius, $R_{0}\approx19\,\text{nm}$.
The effect of tuning the Förster radius through the chemical potential
is evident. In Fig.~\ref{fig:08b} the values of the Förster radius
are smaller for the different values of the chemical potential, $\mu$,
due to the fact the QEs-GM distance is increased. When $\mu=1.0\,\text{eV}$
we have the largest value of the Förster radius, $R_{0}=24.7\,\text{nm}$,
due to our choice of the donor and acceptor. For the off-resonance
case, $\mu=0.4\,\text{eV}$, the Förster radius is approximately the free-space
value, $R_{0}\approx19\,\text{nm}$.

At larger donor-acceptor separations, we use the following expression
to fit the calculated ET rate
\begin{equation}
f(x)=\frac{A}{x}\exp\left(-\frac{2x}{B}\right),\label{eq:16}
\end{equation}
which represents the dependence of the GP field intensity on the in-plane
separation $x$ (the factor 2 in the exponential results from the square
of the Green's tensor, as does the $x$ in the denominator -- the
GP field has a factor of $1/\sqrt{x}$). The fitting parameter $B$
is tabulated in Table~\ref{tab:02}, together with the corresponding
propagation length of the GP along the interface of the GM, $L_{\text{SP}}$.
As is clear, the correspondence between these parameters is very good
indeed, confirming that, away from the near-field, the interaction
between donor and acceptor occurs primarily through the GP excited
by the donor at the surface of the GM. As we increase the distance
between the QEs and the GM, the Förster regime dominates further away
from the near-field, as can be seen from the fact that the intersection
between the two fitting curves moves to larger distances. This is
due to the small value of the penetration depth of the GP, $\delta_{\text{SP}}=6.6\,\text{nm}$
at $\mu=1.0\,\text{eV}$. 
\begin{table}
\caption{\label{tab:02}}

\begin{tabular}{|c|c|c|c|}
\hline 
$z$ (nm) & $\mu$ (eV) & $L_{\text{SP}}$ (nm) & $B$ (nm)\tabularnewline
\hline 
\hline 
5 & 0.8 & 379.23 & 376.82\tabularnewline
\hline 
5 & 1.0 & 890.31 & 884.86\tabularnewline
\hline 
10 & 0.8 & 379.23 & 376.99\tabularnewline
\hline 
10 & 1.0 & 890.31 & 886.47\tabularnewline
\hline 
\end{tabular}
\end{table}
 In the next paragraph we will further consider the influence of the
penetration depth to the ET rate, $k_{\text{ET}}$.

Figs.~\ref{fig:08c} and \ref{fig:08d} consider the ET rate, $k_{\text{ET}}$,
as a function of the donor-acceptor separation, for the case when
the donor position is kept fixed at (\ref{fig:08c}) $z_{\text{D}}=5\,\text{nm}$
and (\ref{fig:08d}) $z_{\text{D}}=10\,\text{nm}$ above the GM, and
the separation between the donor-acceptor, $d=z_{\text{D}}-z_{\text{A}}$
is varied, for $\mu=1.0\,\text{eV}$ and $\mu=0.4\,\text{eV}$. At
small separations we again use the Förster model fitting presented
earlier. To fit the behavior of the ET rate, $k_{\text{ET}}$, below
the GM we choose the expression $f(z)=Ae^{-2z/B}$, where the parameter
$B$ will be connected with the penetration depth of the GP, $\delta_{\text{SP}}$.
In both figures, the GM position is denoted by the dashed vertical
line. In Fig.~\ref{fig:08c}, for which the donor position is very
close to the GM ($z_{\text{D}}=5\,\text{nm}$), the behavior of the
ET rate immediately below the GM is not trivial and comes from various
contributions, such as direct interaction and GP-coupling. On the
other hand, in panel \ref{fig:08d}, for which $z_{\text{D}}=10\,\text{nm}$,
we can use the fitting function $f(z_{\text{A}})$, and find for $B$
the value $B=\delta_{\text{SP}}=6.6\,\text{nm}$, showing that the
main contribution to the ET rate comes from the GP on the GM. For
the $\mu=0.4\,\text{eV}$ case the ET rate, $k_{\text{ET}}$, is almost
uninfluenced by the presence of the GM. The quantity $\mathfrak{D}$
has a maximum value of 0.2 for the donor-GM distance of 5~nm and
$\mu=1.0\,\text{eV}$.

\section{Summary and Conclusions\label{sec:04}}

We have considered in this contribution the behavior of quantum systems
placed near a free-standing graphene monolayer. The graphene monolayer
can support graphene surface plasmon modes, tightly confined to the
surface and having large propagation distances along the graphene
monolayer.

We have begun by investigating the conditions of strong and weak coupling
between a quantum system and the surface plasmon-polariton on the
graphene monolayer. We have seen that for reasonably large values
of the chemical potential $\mu>0.4\,\text{eV}$ and any transition
energies of the QE not in the THz regime, the weak coupling conditions
are fulfilled. We can thus calculate such quantities as spontaneous
emission and energy transfer functions. We have found that both of
these quantities are enhanced, compared to their free space values,
due to efficient coupling to the graphene plasmon modes.

Due to the competition of the donor-acceptor energy transfer process
with other donor decay processes, we have defined the energy transfer
efficiency, $\eta$, and have investigated the influence of the graphene
plasmons on this quantity. We have shown that the energy transfer
efficiency, $\eta$, can reach values above $50\%$ for distances
up to $300\,\text{nm}$ along the graphene monolayer. This process
can be controlled by tuning the value of the chemical potential, e.g.~through
gating.

Finally, we investigate the ET rate, $k_{\text{ET}}$, varying the
donor-acceptor in-plane separation and distance from the GM, for various
values of the chemical potential, $\mu$. The ET rate, when the in-plane
distance between the donor and acceptor is varied, has two major contributions:
the Förster-type mechanism dominates at small separations, while the
GP contribution dominates at large distances. The Förster-type ET
rate follows a $x^{-6}$ dependence, with an increased Förster radius
value, due to the presence of the GM. The Förster radius value is
increased from a free-space value of $R_{0}=19\,\text{nm}$, to $R_{0}=30.7\,\text{nm}$
when $z_{\text{D}}=z_{\text{A}}=5\,\text{nm}$ and $\mu=1.0\,\text{eV}$.
At larger distances, the main contribution comes from the GP propagation;
the transition from the Förster to the GP-propagation mechanism depends
on the distance of the donor-acceptor pair from the GM, and it occurs
at donor-acceptor separations ranging from a few nm to a couple of
tens of nm. When the $z$-distance between donor-acceptor is varied,
for $x=0$, the behavior is somewhat more complicated, but the GP
penetration depth still dictates the interaction length. As the chemical
potential, $\mu$, decreases the ET rate approaches the free space
values. Thus, by varying the value of the chemical potential, we can
switch off the interaction channel or preferentially couple different
species of donor-acceptor resonant pairs of QEs.
\begin{acknowledgments}
This work was supported by the Science Foundation Ireland under grant
No. 10/IN.1/12975. 
\end{acknowledgments}
\appendix

\section{Green's tensor formalism\label{app:A}}

We will consider two planar half-spaces with different dielectric
permittivities, $\varepsilon_{1}$ and $\varepsilon_{2}$. The $z$-direction
is perpendicular to the boundary between the two half-spaces. In order
to calculate the Green's tensor for this system, we use the method
of scattering superposition.\cite{Chew1995}  The Green's tensor
has the form\begin{subequations}\label{eqs:A01} 
\begin{equation}
\mathbf{\mathfrak{G}}^{(11)}(\mathbf{r},\mathbf{s},\omega)=\mathfrak{G}_{h}^{(11)}(\mathbf{r},\mathbf{s},\omega)+\mathfrak{G}_{s}^{(11)}(\mathbf{r},\mathbf{s},\omega),\label{eq:A01a}
\end{equation}
\begin{equation}
\mathbf{\mathfrak{G}}^{(21)}(\mathbf{r},\mathbf{s},\omega)=\mathfrak{G}_{s}^{(21)}(\mathbf{r},\mathbf{s},\omega),\label{eq:A01b}
\end{equation}
\end{subequations}where the first of the two labels in the superscript
$(i1)$ denotes the field point, while the second denotes the source
point. The subscript $s$ denotes the scattering term, always present,
while the homogeneous term $\mathfrak{G}_{h}^{(11)}(\mathbf{r},\mathbf{s},\omega)$
contributes only when the source and field points are in the same
medium.

The scattering terms have the following expression\begin{widetext}\begin{subequations}\label{eqs:A02}

\begin{equation}
\mathfrak{G}_{s}^{(11)}(\mathbf{r},\mathbf{s},\omega)=\frac{i}{8\pi^{2}}\sum_{K}\int\text{d}^{2}k_{\rho}\frac{1}{k_{z1}k_{\rho}^{2}}R_{K}^{+11-}\mathbf{K}(k_{\rho},k_{z1},\mathbf{r})\otimes\mathbf{K}^{*}(k_{\rho},-k_{z1},\mathbf{s})\label{eq:A02a}
\end{equation}
\begin{equation}
\mathfrak{G}_{s}^{(21)}(\mathbf{r},\mathbf{s},\omega)=\frac{i}{8\pi^{2}}\sum_{K}\int\text{d}^{2}k_{\rho}\frac{1}{k_{z1}k_{\rho}^{2}}R_{K}^{-11-}\mathbf{K}(k_{\rho},-k_{z2},\mathbf{r})\otimes\mathbf{K}^{*}(k_{\rho},-k_{z1},\mathbf{s})\label{eq:A02b}
\end{equation}
\end{subequations}where $k_{\rho}=\sqrt{k_{i}^{2}-k_{zi}^{2}}$ is
the in-plane propagation constant, $k_{zi}$ is the perpendicular
propagation constant in medium $i$, and $k_{i}=\frac{\omega}{c}\sqrt{\varepsilon_{i}}$
is the wavenumber in medium $i$ ($i=1,2$). The above expressions
involve a summation over $\mathbf{K}$ which represents $\mathbf{M}$
and $\mathbf{N}$, or the transverse electric (TE) and transverse
magnetic (TM) modes. 

We impose the following continuity conditions at the boundary between
the two half spaces, $z=0$,\begin{subequations}\label{eqs:A03}
\begin{equation}
\hat{z}\times\left[\mathfrak{G}^{(11)}(\mathbf{r},\mathbf{s},\omega)-\mathfrak{G}^{(21)}(\mathbf{r},\mathbf{s},\omega)\right]_{z=0}=0,\label{eq:A03a}
\end{equation}
\begin{equation}
\hat{z}\times\left[\boldsymbol{\nabla}\times\mathfrak{G}^{(11)}(\mathbf{r},\mathbf{s},\omega)-\boldsymbol{\nabla}\times\mathfrak{G}^{(21)}(\mathbf{r},\mathbf{s},\omega)\right]_{z=0}=-i\frac{4\pi}{c}k_{0}\sigma\hat{z}\times\hat{z}\times\mathfrak{G}^{(21)}(\mathbf{r},\mathbf{s},\omega),\label{eq:A03b}
\end{equation}
\end{subequations}\end{widetext}where $\sigma$ is the surface conductivity.

Using Eqs.~(\ref{eqs:A02}) in (\ref{eqs:A03}) we obtain the generalized
Fresnel coefficients, which have the form,\cite{Hanson2013,Nikitin}
\begin{subequations}\label{eqs:A04}
\begin{equation}
R_{M}^{11}=\frac{k_{z1}-k_{z2}-2\alpha k_{0}}{k_{z1}+k_{z2}+2\alpha k_{0}},\quad R_{N}^{11}=\frac{k_{2}^{2}k_{z1}-k_{1}^{2}k_{z2}+2\alpha k_{0}k_{z1}k_{z2}}{k_{2}^{2}k_{z1}+k_{1}^{2}k_{z2}+2\alpha k_{0}k_{z1}k_{z2}}\label{eq:A04a}
\end{equation}
\begin{equation}
R_{M}^{21}=\frac{2k_{z1}}{k_{z1}+k_{z2}+2\alpha k_{0}},\quad R_{N}^{21}=\frac{2k_{1}k_{2}k_{z1}}{k_{2}^{2}k_{z1}+k_{1}^{2}k_{z2}+2\alpha k_{0}k_{z1}k_{z2}},\label{eq:A04b}
\end{equation}
\end{subequations}where $\alpha=2\pi\sigma/c$.

\section{Spontaneous Emission and Energy Transfer Rates\label{app:B}}

The spontaneous emission function $\gamma$ has the form:\cite{Dung1998,Dung2000}
\begin{equation}
\gamma\left(\mathbf{r},\omega\right)=\frac{2\omega^{2}\mu^{2}}{\hbar\varepsilon_{0}c^{2}}\text{Im}\left[\mathbf{n}_{\mu}\cdot\mathfrak{G}(\mathbf{r},\mathbf{r},\omega)\cdot\mathbf{n}_{\mu}\right],\label{eq:B01}
\end{equation}
where $\mathbf{n}_{\mu}$ is a unit vector along the direction of
the transition dipole moment of the emitter, $\mu$, and $\mathfrak{G}\left(\mathbf{r},\mathbf{r},\omega\right)$ is
the Green\textquoteright{}s tensor introduced in Appendix \ref{app:A}.

A useful quantity to introduce is the normalized SE rate, defined
as 
\begin{equation}
\tilde{\gamma}=\frac{\gamma}{\gamma_{0}}=n_{i}+\frac{6\pi c}{\omega}\textrm{Im}\left[\mathbf{n}_{\mu}\cdot\,\mathfrak{G}_{s}(\mathbf{r},\mathbf{r},\omega)\cdot\mathbf{n}_{\mu}\right],\label{eq:B02}
\end{equation}
where the expression for $\gamma_{0}$ is given by the Einstein $A$-coefficient
as $\gamma_{0}=\omega^{3}\mu^{2}/(3\pi\hbar\varepsilon_{0}c^{3})$.
The subscript $s$ on the Green's tensor denotes the scattering part
of this quantity, which we introduced in Eqs.~(\ref{eqs:A01}) and
$n_{i}=\sqrt{\varepsilon_{i}}$ is the refractive index of the medium
into which the quantum system is embedded. The normalized SE rate
gives either an enhancement ($\tilde{\gamma}>1$) or a reduction ($\tilde{\gamma}<1$)
of the SE rate compared to its free-space value, $\gamma_{0}$.

When dealing with statistical ensembles of emitters, the emission
spectrum will be different from that of a single emitter, which we
have taken to have a $\delta$-shape. The emission rate for the ensemble
can then be expressed as:
\begin{equation}
k_{\text{SE}}=\int\limits _{0}^{\infty}\text{d}\lambda f_{D}(\lambda)\gamma(\lambda),\label{eq:B03}
\end{equation}
where $f_{\text{D}}(\lambda)$ is the area-normalized emission spectrum
of the emitter, with $\int\limits _{0}^{\infty}\text{d}\lambda f_{\text{D}}(\lambda)=1$. 

Furthermore, we introduce the ET function, $\Gamma$, between a donor-acceptor
pair, which has the form\cite{Dung2002}

\begin{equation}
\Gamma(\mathbf{r}_{\text{A}},\mathbf{r}_{\text{D}},\omega)=\frac{2\pi}{\hbar^{2}}\left(\frac{\omega^{2}}{c^{2}\varepsilon_{0}}\right)^{2}|\boldsymbol{\mu}_{\text{A}}\cdot\mathfrak{G}(\mathbf{r}_{\text{A}},\mathbf{r}_{\text{D}},\omega)\cdot\boldsymbol{\mu}_{\text{D}}|^{2},\label{eq:B04}
\end{equation}
where again $\mathfrak{G}(\mathbf{r}_{\text{A}},\mathbf{r}_{\text{D}},\omega)$
is the Green's tensor for the particular geometry, $\mathbf{r}_{\text{D(A)}}$
is the position of the donor $D$ (acceptor $A$) and $\mathbf{\boldsymbol{\mu}}_{\text{D(A)}}$
is the transition dipole moment of the donor $D$ (acceptor $A$).
The above expression for the energy transfer function depends on the
donor-acceptor pair through the emission frequency of the donor and
the transition dipole moment of the donor and acceptor. The influence
of the geometry is completely encapsulated in the Green's tensor,
being proportional to the electric field intensity, through the square
of the Green's tensor.

To consider only the influence of the geometry on a general donor-acceptor
pair, we now introduce the normalized ET function for the system,
$\tilde{\Gamma}$, defined as 
\begin{equation}
\tilde{\Gamma}(\omega)=\frac{\Gamma(\omega)}{\Gamma_{0}(\omega)}=\frac{|\mathbf{n}_{\text{A}}\cdot\mathfrak{G}(\mathbf{r}_{\text{A}},\mathbf{r}_{\text{D}},\omega)\cdot\mathbf{n}_{\text{D}}|^{2}}{|\mathbf{n}_{\text{A}}\cdot\mathfrak{G}_{0}(\mathbf{r}_{\text{A}},\mathbf{r}_{\text{D}},\omega)\cdot\mathbf{n}_{\text{D}}|^{2}},\label{eq:B05}
\end{equation}
where $\mathfrak{G}_{0}(\mathbf{r}_{\text{A}},\mathbf{r}_{\text{D}},\omega)$
is the Green's tensor in free space and $\mathbf{n}_{\text{D(A)}}$
is a unit vector in the direction of $\boldsymbol{\mu}_{\text{D(A)}}$. 

Analogously to the case of the SE rate, when considering statistical
ensembles of donors and acceptors, the donor emission spectrum $f_{\text{D}}(\lambda)$
and acceptor absorption cross-section $\sigma_{\text{A}}(\lambda)$
need to be taken into account when calculating the energy transfer
rate. We, therefore, hav:\cite{Zhang2014}
\begin{equation}
k_{\text{ET}}=36\pi^{2}Y_{\text{D}}k_{\text{SE}}\int\limits _{0}^{\infty}\frac{\text{d}\lambda}{\lambda^{2}}f_{\text{D}}\left(\lambda\right)|\mathbf{n}_{\text{A}}\cdot\mathfrak{G}(\mathbf{r}_{\text{A}},\mathbf{r}_{\text{D}},\lambda)\cdot\mathbf{n}_{\text{D}}|^{2}\sigma_{\text{A}}\left(\lambda\right),\label{eq:B06}
\end{equation}
where $Y_{\text{D}}$ is the intrinsic quantum yield of the donor.
We have used this expression to calculate the energy transfer rate
between donors and acceptors with specific emission and absorption
spectra and to investigate how the energy transfer process competes
with the emission process of the donor.

\bibliographystyle{prsty}


\end{document}